\documentclass[aps]{revtex4}

\usepackage{amsmath,amsfonts}
\usepackage{graphicx}
\usepackage{hyperref}

\newcommand{\diag}{\mathop{\text{diag}}}

\begin{document}

\title{Fomin's conception of quantum cosmogenesis}

\author{Marek Szyd{\l}owski}

\affiliation{Astronomical Observatory,
Jagiellonian University, Orla 171, 30-244 Krak{\'o}w, Poland}

\affiliation{Mark Kac Complex System Research Centre, Jagiellonian University, 
Reymonta 4, 30-059 Krak{\'o}w, Poland}

\author{Jacek Golbiak}

\affiliation{Department of Theoretical Physics, 
John Paul II Catholic University of Lublin, Al. Rac{\l}awickie 14, 20-950 Lublin, Poland}

\begin{abstract}
The main aim of this paper is to extend the early approach to quantum 
cosmogenesis provided by Fomin. His approach was developed independently to the 
well-known Tryon description of the creation of the closed universe as a 
process of quantum fluctuation of vacuum. We apply the Fomin concept to derive 
the cosmological observables. We argue that Fomin's idea from his 1973 work, 
in contrast to Tryon's one has impact on the current Universe models and the 
proposed extension of his theory now can be tested by distant supernovae SNIa. 
Fomin's idea of the creation of the Universe is based on the intersection of 
two fundamental theories: general relativity and quantum field theory with the 
contemporary cosmological models with dark energy. As a result of comparison 
with contemporary approaches concerning dark energy, we found out that Fomin's 
idea appears in the context of the present acceleration of the Universe 
explanation: cosmological models with decaying vacuum. Contemporary it appears 
in the form of Ricci scalar dark energy connected with the holographic 
principle. We show also that the Fomin model admits the bounce instead of the 
initial singularity. We demonstrate that the Fomin model of cosmogenesis can 
be falsified and using SNIa data the values of model parameters is in 
agreement with observations.
\end{abstract}

\maketitle

\section{Introduction}

Cosmogony is defined as a study of the origin (cosmogenesis) of the Universe 
in the physical aspect. It asks the question how the Universe came into being. 
By comparing this process with human life, we called the birth of the Universe 
cosmogenesis \cite[p.515]{Harrison:2000cs}. Among different conceptions of 
cosmogenesis which offer the possibility of describing the origin of the 
Universe as a physical process, there are cosmogenic theories of the 
spontaneous creation from vacuum instability. According to these theories, the 
Universe was created in a spontaneous way, from a quantum fluctuation. Edward 
Tryon \cite{Tryon:1973iu,Tryon:1984wm} is usually considered as the scientist 
who first supported such theories in 1973 by arguing that total mechanical 
energy of the closed Universe is zero and effects of quantum fluctuation of 
the vacuum were important when the Universe was coming to the existence.

The second approach developed independently by Fomin 
\cite{Fomin:1973gi,Fomin:1975gi} in 1973\footnote{%
Preprint titled ``Gravitational Instability of Vacuum and Cosmological 
Constant'' was reported at a seminar on January 26, 1973 (this date was 
written on the abstract page). This paper was issued by Institute of 
Theoretical Physics in Kiev as preprint no. ITP-73-137R \cite{Fomin:1973gi}, 
but it was withdrawn from circulation. The reason was a reference to Andrei 
Sakharov's paper, who was sent to exile in the same year, and a director of 
the institute had to be requested to remove, if possible, the name of Sakharov 
from all the reference lists. One can suppose it was rather not the matter of 
principle but to avoid some troubles. The paper was then published with the 
replaced reference (Problems of modern cosmology, ed. by W.~A. Ambartsumyan) 
two years later \cite{Fomin:1975gi}.}
was based on the idea that the closed Universe originated from gravitational 
instability of the vacuum is realized in the background of general relativity 
theory, rather than on Newtonian one, like in Tryon's paper. Fomin's 
contribution is only addressed to in a footnote in Vilenkin's paper on 
quantum tunneling \cite{Vilenkin:1984qc}.

The main aim of this paper is to show that Fomin's idea offers deeper 
understanding of the origin of the Universe from the vacuum fluctuation for 
two reasons: the mechanism proposed by Fomin is based on Einstein general 
relativity in contrast to Tryon's idea formulated on the background of the 
Newtonian theory of gravity. Moreover, Tryon's idea cannot concern the curved 
closed universe. Nevertheless, the Newtonian theory can be useful in the 
context of discovery, but of course, it does not describe the early epoch of 
the universe when both gravitational and quantum effects plays a crucial 
role.

We can discover a new contemporary context for Fomin's theory by considering 
the cosmological models in which his idea is realized, providing the 
conservation condition is postulated. We find two basic categories of 
the cosmological models in which the universe is generally relativistic and 
originated from quantum fluctuation via Fomin's proposition. We also 
demonstrate that in contrast to Tryon's model, the models obtained in this 
paper can be confronted with modern cosmological observations of the distant 
supernovae SNIa. The critical analysis of Tryon's paper is not the subject of 
the present paper. Some interesting remarks and critiques of Tryon's idea can 
be found in McCabe's papers \cite{McCabe:2005gu,McCabe:2005gv}. We share 
McCabe's comments on Tryon's paper in many points but the correct definition of 
the energy seems to be crucial for the investigation of the creation of the 
Universe. Many authors have concluded recently that energy of the flat and 
closed FRW universes are equal to zero locally and globally. Unfortunately such 
conclusions originate from coordinate dependent calculations which are 
performed in the special comoving coordinates (called the Cartesian 
coordinates), by using energy momentum tensor of matter and gravity 
\cite{Garecki:2005vf}. 

However, let us note that all expressions for the conserved energy-momentum 
tensor in general relativity can be written as a divergence of a 
superpotential. The integral over space can then be expressed as a surface 
integral over the boundary, which is zero for the closed Universe. This 
conclusion does not depend on which metric or which coordinate system is used 

The organization of this paper is as follows. In Section 2 we present Fomin's 
idea in details. Section 3 is devoted to the construction of cosmological 
models from basic Fomin's assumptions --- first principles. In Section 4 these 
cosmological models are tested by the data from observation of distant 
supernovae type Ia, which offer the possibility of explaining the present 
acceleration of the Universe.

\section{Fomin's idea of the origin of the Universe from quantum fluctuations}

Fomin's main goal (like Tryon's) was to describe the origin of the Universe as 
a whole without violation of any conservation laws. Since at that moment gravitational 
fields were very strong, the gravitational interaction of the primordial vacuum 
with gravity should be taken into account. In both Tryon's and Fomin's papers, 
the vacuum is chosen as an initial state of the Universe (a metagalaxy in 
Fomin's terminology). Note that if strong gravitational fields are important, 
then effects of general relativity (not special relativity) should be included 
at the very beginning. This is the very reason why Fomin formulated the problem 
in the context of general relativity. In the introduction to his paper we can 
find an analogy to the Newtonian cosmology, but he noted the absence of the 
kinetic term $(\dot{a}^2/2)$ in the energy balance equation $Mc^2+V(M,a)=0$
where $M$ is the mass of the universe particle and $V(M,a)$ is the potential 
energy. It would be worth mentioning at this point that the notion of energy 
for the FRW models is well defined from the physical point of view because we 
are not dealing with isolated systems which are asymptotically flat spacetimes. 
Note that formally, from Pirani's or Komar's formula we can obtain zero energy 
for the Friedmann-Robertson-Walker (FRW) models but there is no time Killing 
vector and we have geodesic congruence of privileged observers. That is why, in 
our opinion, this result does not make much physical sense. Moreover, the 
result of energy calculations following Pirani's or Komar's formula depends on 
the choice of a coordinate system.

In the introduction to Fomin's paper we find remarks that the process of 
quantum cosmogenesis should be considered in the presence of a strong 
gravitational field and therefore the effects of general relativity should be 
included at the very beginning. In contrast to Tryon, Fomin considered an 
evolving non-static universe. It is well known that the FRW dynamics can be 
represented by the notion of a particle of unit mass moving in the potential, 
following an equation analogous to the Newtonian equation of motion, namely
\begin{subequations}
\label{eq:1}
\begin{align}
\mathcal{H} &= \frac{\dot{a}^2}{2} + V(a) \\
\ddot{a} &= -\frac{\partial V(a)}{\partial a}
\end{align}
\end{subequations}
where $p_{a} = \dot{a}$ and $a$ are generalized momenta and coordinates, 
respectively, $H$ is the Hamiltonian and $V$ is the potential function of the 
scale factor $a$, if $V$ is a decreasing function of $a$, the Universe is 
accelerating \cite{Szydlowski:2003fg,Szydlowski:2003cf}.

System (\ref{eq:1}) should be considered on the energy level $H=E=-\frac{K}{2}$ 
where $K=0,\pm 1$ is the curvature constant (we use here and in next sections 
the system of units in which $ 8\pi G =c= 1$. Therefore, only for a flat 
relativistic system we have the energy level $E$ equals to zero. It is 
interesting that the FRW model based on Fomin's conception can be formulated in 
terms of Hamiltonian systems of a Newtonian type in the form (\ref{eq:1}) with 
the corresponding function of the scale factor $V(a)$.

\begin{figure}
\includegraphics[width=0.8\textwidth]{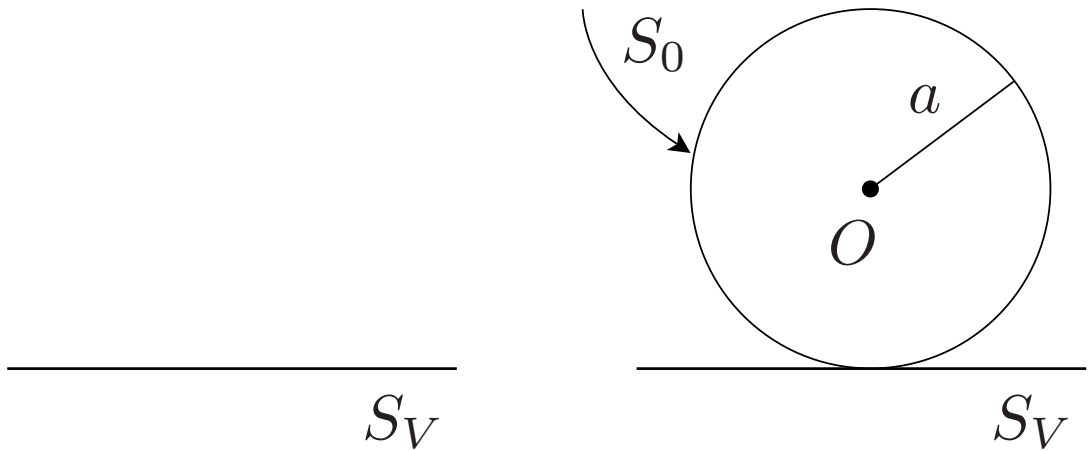}
\caption{Illustration of emerging of a closed null spacetime system $S_0$ from 
vacuum $S_V$ (the idea of picture comes from Fomin's papers 
\cite{Fomin:1973gi,Fomin:1975gi}).}
\label{fig:1}
\end{figure}

In Fomin's scenario, the Universe emerged from a physical state, which he 
called the ``null vacuum'' system $S_V$. Vacuum is assumed as an initial state 
of further evolution of the Universe. This system produced particles and 
antiparticles in a spontaneous manner without violating any conservation laws. 
Then emerging matter of mass $m$, (called bimatter by Fomin), gives rise to the 
growth of entropy. Fomin argued that such a process of decaying vacuum is 
energetically preferred and both entropy and the number of particles and 
antiparticles will be growing at the cosmological time. The growth of the 
number of particles is strictly related to the size of the system, as we assume 
that the created Universe is closed (see Fig.~\ref{fig:1}).

Fomin pointed out that for the description of the contribution of decaying 
vacuum (in a phenomenological way), the Einstein field equations should be 
generalized. Hence the vacuum is characterized by an energy momentum tensor of 
the form
\begin{equation}
\label{eq:2}
T_{\mu \nu}^{\text{vacuum}} = - \lambda(x) g_{\mu \nu}
\end{equation}
where the energy density of vacuum $\lambda$ depends on the scalar curvature of 
the spacetime
\begin{equation}
\label{eq:3}
\lambda(x) = k \kappa^{-1}R(x)
\end{equation}
where $\kappa = 8\pi G/c^4$ and $k$ is a dimensionless parameter. In Fomin's 
paper $\lambda$ was assumed to be positive (however we will not restrict it 
in the further sections). There are some physical reasons for such a choice of 
the parameterization by considering a heuristic argument based on the 
conjecture about quantum gravity. Let us suppose that $\rho_{\lambda}$ raises 
due to a quantum gravity fluctuation. Then the action 
$S_{g} \propto \int d^{4}x \, \sqrt{-g}R$ might be nonzero due to quantum 
$\Delta V \Delta t$. So $\Delta D \simeq 1$ \cite{Garay:1994en}. Therefore 
$\Delta S \propto \frac{\Delta E}{\Delta V} \propto R$ 
due to energy-time uncertainty relation $\Delta E \Delta t \sim 1$.

\section{Fomin's model re{\"e}xamined}

Let us assume that a quantum process proposed by Fomin can be modelled, in a 
phenomenological way, by the energy momentum tensor in the form (\ref{eq:2}) 
and when the source of gravity is a perfect fluid with energy density $\rho$ 
and pressure $p$. Therefore the Einstein equation assumes the form
\begin{equation}
\label{eq:4}
R_{\mu \nu} - \frac{1}{2} R g_{\mu \nu} = \kappa \bar{T}_{\mu \nu}
\end{equation}
where $\mu \nu = 0, 1, 2, 3$ and energy momentum tensor is
\begin{equation}
\label{eq:5}
\bar{T}_{\mu \nu} = T_{\mu \nu} - \lambda g_{\mu \nu} 
= (\rho + p) u_{\mu} u_{\nu} + (p - \lambda(x))g_{\mu \nu}
\end{equation}
\[
g = \diag(-1,1,1,1), u_{\mu}u^{\mu} = -1, u^{\mu} = (1,0,0,0), u_{\mu}=(-1,0,0,0)
\]
where $u=(u_0,u_i)$ is a four-vector of velocity.

The form of the Einstein equation (\ref{eq:4}) is more suitable for our aims 
because the contribution from the energy density of the vacuum was shifted to 
the energy momentum tensor. It is clear from divergence of the field equations 
\begin{equation}
\label{eq:6}
\bar{T}^{\mu \nu}_{\ \ \ ;\nu} = \left[ R^{\mu \nu} - \frac{1}{2} R g^{\mu \nu} \right]_{;\nu} = 0.
\end{equation}
Hence the energy-momentum tensor is given in the diagonal form
\begin{equation}
\label{eq:7}
\bar{T}^{\mu}_{\nu} = (\rho_{\text{eff}}+p_{\text{eff}})u_{\mu} u_{\nu} 
+ p_{\text{eff}} g_{\mu \nu}
\end{equation}
where $p_{\text{eff}} = p - \lambda$, $\rho_{\text{eff}} = \rho + \lambda$ 
and $\lambda(t) = k \kappa^{-1} R$, $R=\frac{1}{4k-1}\kappa T$, $T = T^{\mu}_{\ \mu} = 3p - \rho$ 
which is obtained following Fomin's assumption from the basic equation
\begin{equation}
\label{eq:11}
R_{\mu \nu} - \left( \frac{1}{2} - k\right) R g_{\mu \nu} = T_{\mu \nu}
\end{equation}
with the natural system of units $8 \pi G = c = 1$.
 
Since we are going to construct the cosmological model, as the first 
approximation, we consider the model, which is spatially homogeneous and 
isotropic, therefore its spacetime metric is given in the form
\begin{equation}
\label{eq:8}
ds^2 = dt^2 - a^{2}(t) [d\chi^2 + \sin^{2}\chi (d\theta^{2} + \sin^{2} 
\theta d\phi^{2})]
\end{equation}
where $t$ is the cosmological time; $a(t)$ is a scale factor; $r$, $\theta$, 
$\phi$ are standard spherical coordinates; $\chi \colon \chi = \arcsin r$,
$d\chi = \frac{dr}{\sqrt{1-r^2}}$, because it is assumed a positive value of 
curvature spatial slices of constant time of hypersurfaces.

In Fomin's original paper, the evolution of the spacetime is governed by a 
modified Einstein equation rather than the modified energy-momentum tensor 
(\ref{eq:7}). In our extension of Fomin's model we shift the contribution 
coming vacuum instability to the energy-momentum tensor. While both 
formulations are formally equivalent (gives the same formula $H(z)$), the 
presented approach offers the possibility of adding the conservation condition 
to the basic equation, which makes the model in a closed form. Due to 
homogeneity and isotropy we have $\lambda(x)=\lambda(t)$ and dynamics is 
equivalent to the dynamics of FRW cosmological models with some effective 
energy density $\rho_{\text{eff}}$ and pressure $p_{\text{eff}}$
\begin{align}
\label{eq:9}
\rho_{\text{eff}} &= \rho + \lambda(x) = \rho + \frac{k}{4k - 1}(3p - \rho) \\
\label{eq:10}
p_{\text{eff}} &= p -\lambda(x) = p - \frac{k}{4k - 1}(3p - \rho) \\
\lambda(x) &= \frac{k}{4k-1}(3p-\rho).
\end{align}

Note that if we treat $\rho_{\lambda} = \lambda(x)$ as an energy density of 
dark energy then
\[
\frac{\rho_{\lambda}}{\rho} = \frac{\Omega_{\lambda}}{\Omega_{\text{m}}} 
= \frac{k}{1- 4k} = \text{const}.
\]
This means that so-called coincidence problem (i.e. why dark matter and 
dark energy are of the same order today) is solved naturally in the framework 
of the Fomin model. If we require 
\[
\frac{\Omega_{\lambda}}{\Omega_{\text{m}}} \simeq \frac{0.72}{0.28}=\frac{k}{1-4k}
\]
then it is satisfied for $k=0.228$.

Therefore if we choose the form of the equation of state for matter $p=p(\rho)$ 
then one can obtain effective values of pressure and energy density. 
As a result we obtain that the Fomin model belongs to a category of kinematical 
model with a dynamic value of cosmological constant. Obviously, a solution 
with dynamic cosmological constant is possible only if $T^{\mu \nu} \neq 0$ 
(and $T^{\mu \nu}_{\ \ \ ; \nu} \neq 0$). In the absence of matter $(\rho=p=0)$, 
$\lambda$ has got to remain a constant. \cite{Vishwakarma:2002eka} denoted 
that this category models which are invoked to solve a cosmological constant 
problem are in fact consistent with Mach's ideas because an empty spacetime
cannot be a solution of general relativity with the dynamic cosmological 
constant ($\lambda(t)$ in our case). It is a consequence of the fact that a 
conserved quantity is of usual matter and vacuum (and not these two separately) 
following $\bar{T}^{\mu \nu}_{\ \ \ ; \nu} = 0$ which reduces to the following 
equation
\begin{equation}
\label{eq:13}
\frac{d}{dt}\left( \rho_{\text{eff}} a^{3} \right) + p_{\text{eff}} 
\frac{d}{dt} a^{3} = 0
\end{equation}
or
\begin{equation}
\label{eq:14}
\left[ p \frac{d}{dt} a^{3} + \frac{d}{dt} \left( \rho a^{3} \right) \right] 
+ \left[ \frac{d}{dt} \left( \lambda a^{3} \right) - \lambda \frac{d}{dt} 
a^{3} \right] = 0.
\end{equation}
Among the dynamical models of the cosmological constant (for review see 
\cite{Overduin:1998zv,Alam:2004ip,Babic:2004ev,Ray:2004nq}) there are different 
choices of cosmological constant parameterization. In the Fomin model the 
dependence of $\lambda(a(t))$ is determined by the form of the equation of 
state $p=p(\rho)$ through equation (\ref{eq:14}). Some different constraints 
on a decaying cosmological term from the astronomical observation have been 
found, e.g. \cite{Nakamura:2006ij}. As a consequence the Fomin model with the 
decaying term $\lambda(a)$ can be tested through the different astronomical 
observations like distant type Ia supernovae data and Wilkinson Microwave 
Anisotropy Probe (WMAP) data as well as compared with the concordance 
$\Lambda$CDM model. One can say that the Fomin model is a prototype of the 
model that incorporates a cosmic time variation of the cosmological term which 
plays a crucial role in the mechanism of creation of the universe from the 
vacuum. If $\lambda(a)$ is positive then $k \in (0, 1/4)$.

The evolution of the universe is governed by two basic equations which constitute 
the closed system of basic equations for the FRW cosmology (with incorporated 
Fomin's mechanism of gravitational instability of vacuum) when we postulate 
dependence of energy density and pressure on the scalar factor 
($\rho = \rho(a(t)), \quad p = p(a(t)), \quad \lambda = \lambda(a(t))$)
\begin{equation}
\label{eq:16}
\frac{\ddot{a}}{a} = - \frac{1}{6} \left( \rho_{\text{eff}} + 3 p_{\text{eff}} \right) 
= - \frac{1}{6} \left( \rho +3p \right) + \frac{\lambda}{3} 
= - \frac{1}{6} \left( \rho +3p \right) + \frac{k}{3(4k-1)}(3p - \rho).
\end{equation}
and the conservation condition (\ref{eq:14}). To integrate (\ref{eq:14}) and 
(\ref{eq:16}) the form of $p = p(\rho)$ should be postulated. Hereafter 
for simplicity of presentation we assume that 
\begin{equation}
\label{eq:15}
p = \gamma \rho
\end{equation}
where $\gamma = \text{const}$.

Of course equation (\ref{eq:16}) possesses the first integral (called the 
Friedmann first integral)
\begin{equation}
\label{eq:17}
\rho_{\text{eff}} - \frac{3K}{a^{2}} = 3 \frac{\dot{a}^{2}}{a^{2}}
\end{equation}
where $K \in \{ 0, \pm 1 \}$ is a curvature constant.
One can simply check this by differentiation of both sides of equation 
(\ref{eq:17}) with respect of cosmological time and substitution (\ref{eq:17}).

The model under consideration contains a new term in comparison with the 
standard FRW perfect fluid cosmology which is related to the existence of the 
parameter $k$ not equal zero. This parameter measures the effectiveness 
of the process of creation of the Universe from the null system. Note that in 
the radiation epoch, when $p = \rho/3$, the effects of gravitational 
instability of vacuum vanish. We assume following Fomin that quantum effects 
of creation of the Universe are manifested at the epoch in which $\rho -3p$ 
is positive (this condition is distinguished from Fomin's condition of 
positiveness of $\lambda(a)$. Therefore, if $0 \le k \le \frac{1}{4}$, this 
guarantees the positiveness of $\lambda(a(t))$. Fomin pointed out that effect 
of dissociation of vacuum originates in the nonhomogeneous region where 
$\rho - 3p$. During evolution of the Universe, the effects of vacuum 
instability stay small because of local expansion of the space.

With the assumption (\ref{eq:15}) the solution of equation (\ref{eq:16}) takes 
the form
\begin{align}
\label{eq:19}
\rho_{\text{eff}} &= \frac{3k(\gamma + 1) - 1}{4k - 1} \rho \\
\label{eq:20}
p_{\text{eff}} &= \frac{k(\gamma + 1) - \gamma}{4k - 1} \rho 
\end{align}
From (\ref{eq:14}) we obtain as a solution
\begin{equation}
\rho_{\text{eff}} = \rho_{0} a^{-3(1+w_{\text{eff}})} 
= \rho_{0} a^{-3\frac{4k(\gamma + 1) + (k-1)\gamma - 1}{3k(\gamma+1)-1}}.
\end{equation}
(In the special case of dust matter we obtain 
$\rho_{\text{eff}} = \rho_{0} a^{-3\frac{4k-1}{3k-1}}$.)

Hence we determine the equation of state coefficient 
\begin{equation}
\label{eq:21}
w_{\text{eff}} \equiv \frac{p_{\text{eff}}}{\rho_{\text{eff}}} 
= \frac{k(\gamma+1) - \gamma}{3k(\gamma+1) - 1}.
\end{equation}
We obtain from formula (\ref{eq:21}) that the Universe is accelerating for 
dust filled matter if the parameter $k$ belongs to the interval 
$k \in \left( \frac{1}{6}, \frac{1}{3} \right)$; $w_{\text{eff}} < - \frac{1}{3}$.

For our further analysis of constraints from SNIa observational data it is 
useful to represent an evolutional scenario of models in terms of the Hubble 
function and dimensionless density parameters. For the two distinguished cases 
we obtain
\begin{equation}
\label{eq:22}
H^{2} = \frac{\rho_{\text{eff}}}{3} - \frac{1}{a^{2}} = 
\frac{\rho_{\text{eff}}(z)}{3} - \frac{1}{(1+z)^{2}}
\end{equation}
where $1+z=\frac{a_{0}}{a}$; $a_{0} = 1$ is the present value of the scale 
factor and
\begin{equation*}
\frac{H^{2}}{H_{0}^{2}} = \Omega_{K,0}(1+z)^{2} 
+ \Omega_{\text{Card},0} (1+z)^{3\frac{4k(\gamma+1) - \gamma - 1}{3k(\gamma+1)-1}}
\end{equation*}
where $\Omega_{i,0} = \frac{\rho_{i,0}}{3H_{0}^{2}}$ are the density 
parameters for the $i$-th component of the fluid. Of course 
$\sum_{i} \Omega_{i,0} = 1$ from condition $H(z=0)=H_{0}$. 
In the special case of dust matter we obtain the useful relation in the 
context of constraining the model parameters against the astronomical data
\begin{equation}
\label{eq:23}
\left( \frac{H}{H_0} \right)^{2} = \Omega_{K,0}(1+z)^{2} 
+ \Omega_{\text{Card},0} (1+z)^{3\frac{4k-1}{3k-1}}
\end{equation}
where $\Omega_{K,0} + \Omega_{\text{Card},0} = 1$.

We obtain that universe is still accelerating if $w_{\text{eff}} < -1/3$ 
and $\rho_{\text{eff}} > 0$, i.e.
\[
\frac{6k(\gamma +1) - 3\gamma -1}{3k(\gamma +1) -1} < 0 
\qquad \text{and} \qquad 
\frac{3k(\gamma +1) -1}{4k -1} > 0.
\]
For dust it means that that
\[
\frac{6k-1}{3k-1} < 0 \qquad \text{and} \qquad \frac{3k-1}{4k-1} > 0
\]
i.e. that $k \in (\frac{1}{6},\frac{1}{4})$.
Therefore the Fomin model predicts acceleration of the universe driven by 
the mechanism of origin of the universe from gravitational instability of 
vacuum.

In the next section we concentrate on the constraints from recent SNIa 
measurements of the effects arising from possible Fomin's quantum creation 
mechanism which can be phenomenologically manifested by dark energy through 
a modified Friedmann equation.

\section{Distant type Ia supernovae as cosmological probes of quantum cosmogenesis}

Today cosmology appears to be one of the fastest growing parts of physics due 
to new experiments from measurements of the cosmic microwave background 
radiation to measurements of the apparent magnitudes of several high redshift 
supernovae of type Ia, published recently by Riess et al. \cite{Riess:2004nr}. 
For distant supernovae, one can directly observe the apparent magnitude (i.e., 
log of flux $F$) and its redshift. As the absolute magnitude $M$ of any 
supernovae is related to the present luminosity $L$, then relation 
$F=\frac{L}{4\pi d_{L}^{2}}$ can be written as
\begin{equation}
\label{eq:24}
m-M = 5 \log_{10} \left( \frac{d_{L}}{\text{Mpc}} \right) + 25. 
\end{equation}
Usually instead of $d_L$ dimensionless quantity $\frac{H_{0} d_{L}(z)}{c}$ 
is used, and then equation (\ref{eq:23}) changes to
\begin{equation}
\label{eq:25}
m(z) = \bar{M} + 5 \log_{10} \left( \frac{H_{0} d_{L}(z)}{c} \right)
\end{equation}
where the parameter $\bar{M}$ is related to $M$ by the relation
\begin{equation}
\label{eq:26}
\bar{M} = M + 25 + 5 \log_{10} \left( \frac{c H_{0}^{-1}}{\text{Mpc}} \right).
\end{equation}
We know the absolute magnitude of SNIa from its light curve. Therefore we can 
obtain $d_L$ for these supernovae as a function of redshift because the 
apparent magnitude $M$ can be determined from low $z$ apparent magnitude. 
Finally, it is possible to probe dark energy which constitutes main 
contribution to the matter content from $d_{L}(z)$. In our further analysis of 
SNIa data, we estimate models with value of $M \cong 15.955$ determined from 
low redshift relation ($d_{L}(z)$ is then linear) without any prior assumption 
on $H_{0}$. 

We use the standard statistical approach to obtain the best fitting model 
minimizing the $\chi^2$ function. This analysis is supplemented by the maximum 
likelihood method to find confidence ranges for the estimated model parameters.

It is assumed that supernovae measurements came with uncorrelated Gaussian 
errors and the likelihood function $\mathcal{L}$ could be determined from the 
$\chi^{2}$ statistic $\mathcal{L} \propto \exp\left( -\frac{\chi^{2}}{2}\right)$.
In our analysis the latest compilation of SNIa prepared by Riess et al. is 
used (the Gold sample) \cite{Riess:2004nr}. 

Recent measurements of type Ia supernovae as well as other WMAP and 
extragalactic observations, suggest that the expansion of the Universe is an 
accelerating expansion phase. There are different attempts to explain the 
present acceleration of the Universe. Dark energy of unknown form has usually 
been invoked as the most feasible mechanism. Also effects arising from exotic 
physics, like extra dimensions, some modification of the FRW equation can also 
mimic dark energy through a modified Friedmann equation.

It would be interesting to check the consistency of Fomin's 
model with SNIa observations. We begin by evaluating the luminosity distance 
as a function of redshift $z$ as well as the parameters of the model. We 
define the redshift dependence of $H$ as $H(z) = H_{0} E(z)$. For 
Cardassian stylization of Fomin's model of the closed universe with matter 
(baryonic and cold matter) we get that the luminosity distance is given by 
\begin{equation}
\label{eq:27}
d_{L}(z, \Omega_{\text{m},0}, H_{0}) 
= \frac{c(1+z)}{H_{0}} \int_{0}^{z} \frac{dz'}{E(z')}.
\end{equation}

The standard measure of acceleration in cosmology is the dimensionless 
deceleration parameter $q$ defined as 
\begin{equation}
\label{eq:28}
q = - \frac{a \ddot{a}}{\dot{a}^{2}}.
\end{equation}
We can calculate the value of this parameter for both classes of models under 
consideration. Then we obtain
\begin{equation}
\label{eq:29}
q_{0} = \frac{1}{2} (1+3w_{\text{eff}}) 
= \frac{4k(\gamma+1)-\gamma-1}{2[3k(\gamma+1)-1]}.
\end{equation}

We estimate the two-parameter model in the following form
\begin{equation}
\left( \frac{H}{H_0} \right)^2 
= (1-\Omega_{\text{card,0}})(1+z)^2 + \Omega_{\text{Card,0}}(1+z)^{3n}
\end{equation}
where $\Omega_{\text{card},0}$ is positive and less than one, and the parameter 
$n$ is positive. The latter parameter is related to the searched parameter $k$
\begin{equation}
n=\frac{4k-1}{3k -1}
\end{equation}
where we assume $k$ to be $(0,1/4)$.

\begin{figure}
\includegraphics[width=0.45\textwidth]{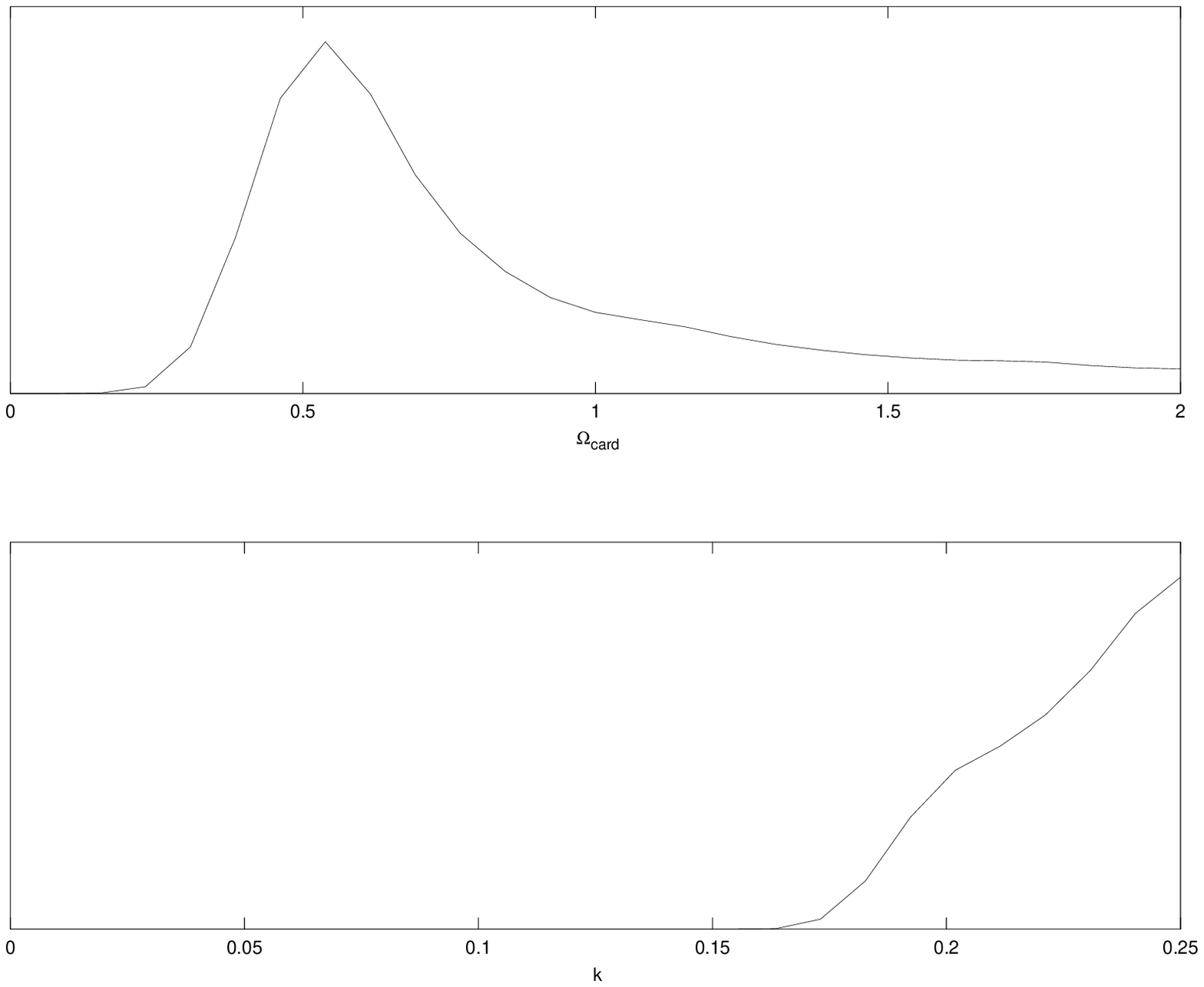}
\includegraphics[width=0.45\textwidth]{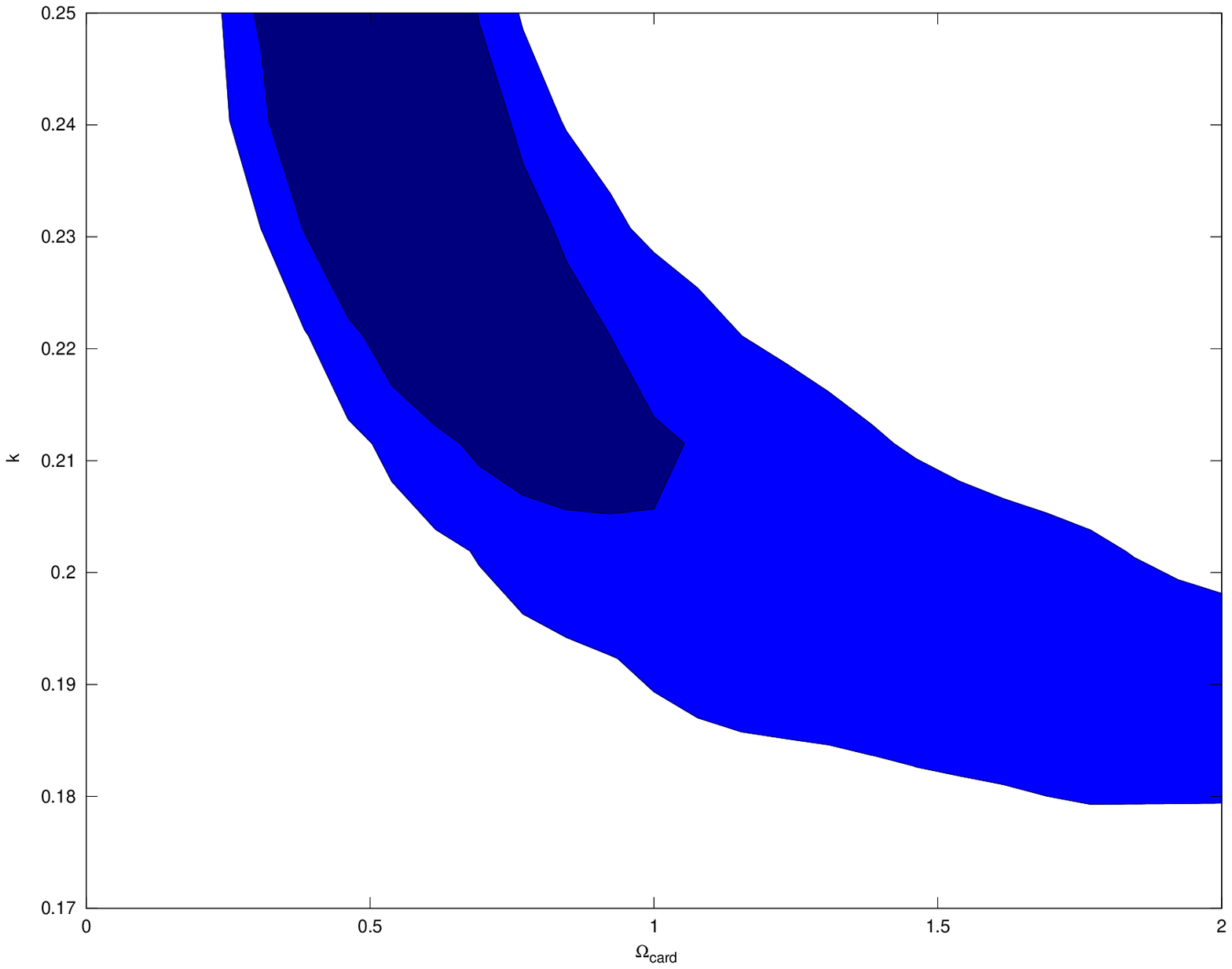}
\caption{The estimation of parameters $\Omega_{\text{Card},0}$ and $k$ for 
the model with $\Omega_{\text{Card},0} \in (0,2)$:
left panel -- posteriori probability distribution functions;
right panel -- posteriori probability confidence levels $68\%$ and $95\%$.}
\label{fig:2}
\end{figure}

\begin{figure}
\includegraphics[width=0.45\textwidth]{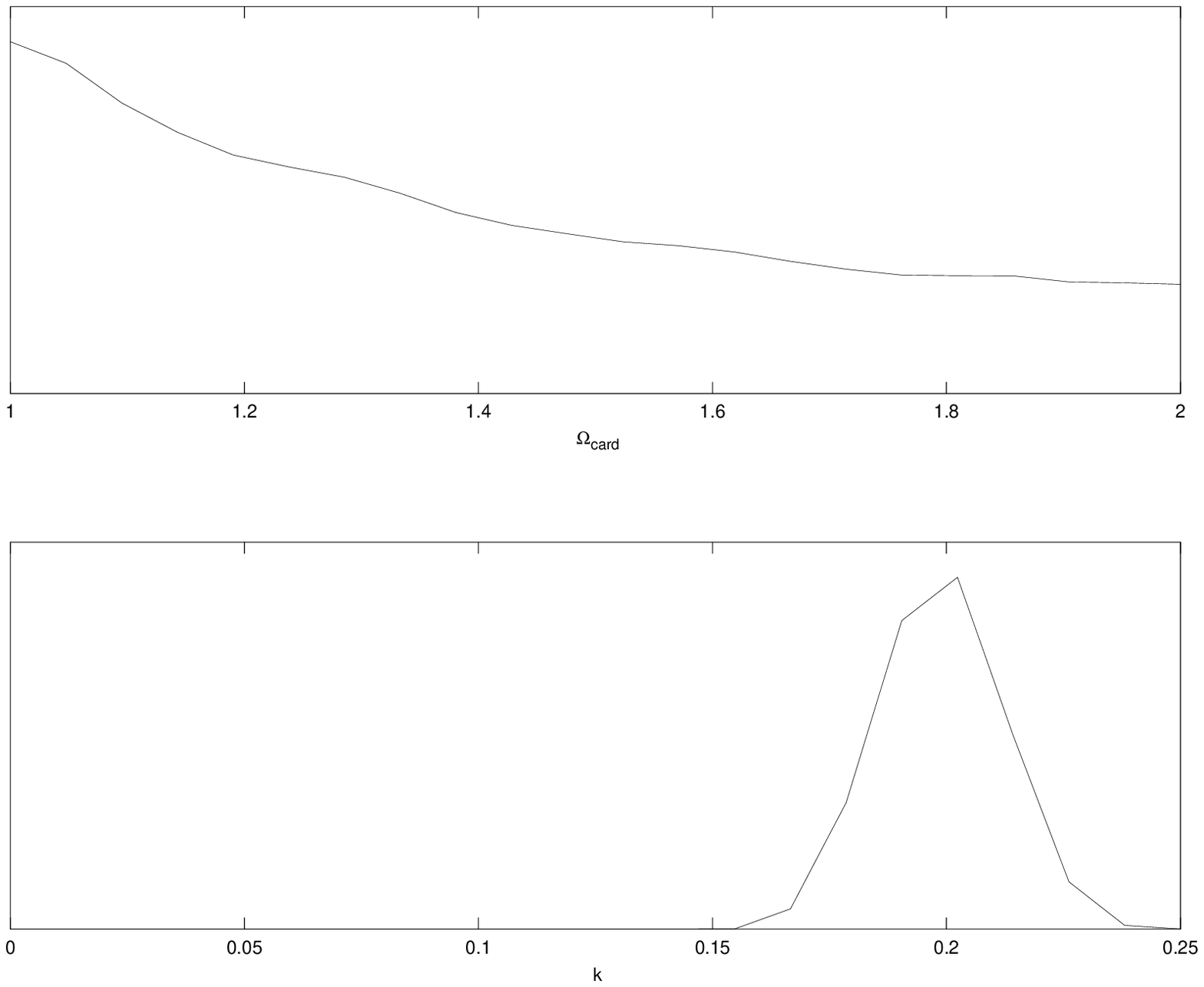}
\includegraphics[width=0.45\textwidth]{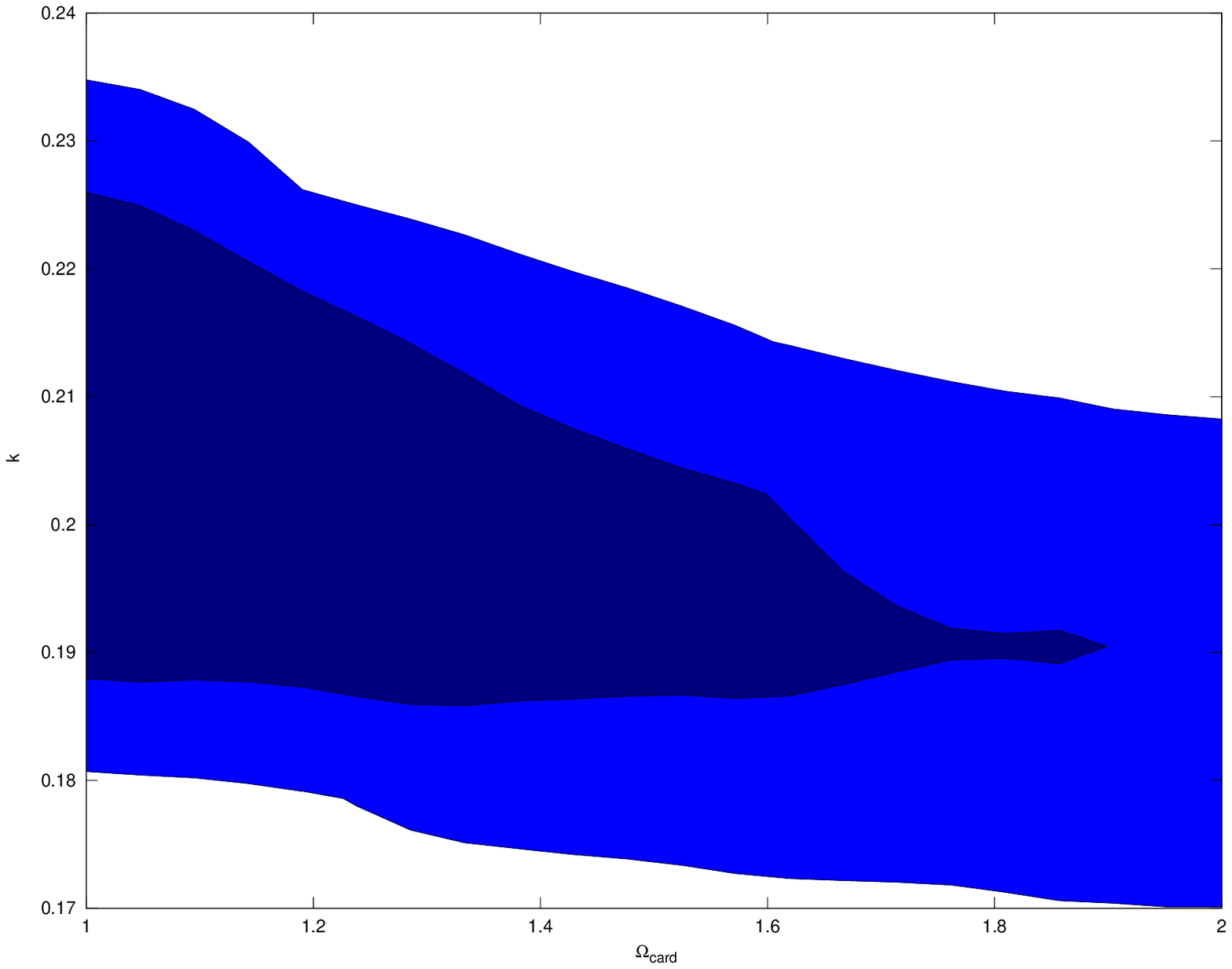}
\caption{The estimation of parameters $\Omega_{\text{Card},0}$ and $k$ for
the model with $\Omega_{\text{Card},0} \in (1,2)$: 
left panel -- posteriori probability distribution functions;
right panel -- posteriori probability confidence levels $68\%$ and $95\%$.}
\label{fig:3}
\end{figure}

We estimate two cases where different regions for $\Omega_{\text{Card},0}$ 
are assumed. The first model corresponds to the case when the curvature of the 
space is undetermined for which $\Omega_{\text{Card},0} \in (0,2)$. The 
second model exactly corresponds to the original Fomin model with the positive 
curvature assumed at the very beginning for which $\Omega_{\text{Card},0} \in 
(1,2)$. In both cases we assume that $k \in (0, 0.25)$. Using the SNIa Union 
sample we estimate both parameters $\Omega_{\text{Card},0}$ and $k$ to find 
the best fit \\
--- for the first model
\[
\Omega_{\text{Card},0}=0.44 \qquad \text{and} \qquad k=0.24 \qquad (\chi^2 = 315.46)
\]
--- and for the Fomin model
\[
\Omega_{\text{Card},0}=1.03 \qquad \text{and} \qquad k=0.21 \qquad (\chi^2 = 318.84).
\]
The expected value of estimated parameters with a $68\%$ confidence interval \\
--- for the first model
\[
\Omega_{\text{Card},0}=0.80^{+0.40}_{-0.32} \qquad \text{and} \qquad k=0.22^{+0.02}_{-0.02},
\]
--- and for the Fomin model
\[
\Omega_{\text{Card},0}=1.40^{+0.34}_{-0.32} \qquad \text{and} \qquad k=0.20^{+0.01}_{-0.01}
\]
These posteriori probability distribution functions are shown in 
Fig.~\ref{fig:2} for the first model and in Fig.~\ref{fig:3} for the Fomin
model. 

The comparison these two models with the $\Lambda$CDM model gives that there 
is the moderate evidence in favour of the $\Lambda$CDM model in both cases. 
This conclusion is based on the Bayes factors analysis. For the first model 
with respect to the $\Lambda$CDM model we obtain the Bayes factor 
$B_{\text{model 1}, \Lambda \text{CDM}} = 3.78^{+0.1}_{-0.1}$. For the second 
model with respect to the $\Lambda$CDM model we obtain the Bayes factor 
$B_{\text{Fomin}, \Lambda \text{CDM}} = 4.59^{+0.13}_{-0.13}$. Our estimation 
shows that the Fomin model is consistent with observations of SNIa.

\section{Dynamics of the cosmological models with Fomin's mechanism creation 
of the Universe}

The dynamics of the model can be represented with the 
help of a particle-like description, namely it can be reduced to the form of 
a particle of unit mass moving in a potential well. The forms of potential 
function for dust matter and matter with $p=\gamma\rho$ models respectively are
\begin{align}
\label{eq:31}
V(a) &= - \frac{1}{2} \left( \Omega_{K,0} + \Omega_{\text{m},0} a^{-3\frac{4k-1}{3k-1} + 2} \right) \\
V(a) &= - \frac{1}{2} \left( \Omega_{K,0} + \Omega_{\text{m},0} a^{-3\frac{4k(\gamma+1)-\gamma -1}{3k(\gamma+1)-1} +2} \right).
\end{align}

There are two different scenarios of an initial state of the Universe namely a 
singularity or a bounce. It depends on the exponent in the potential form 
(\ref{eq:31}). Note that if $-\frac{3(4k-1)}{3k-1} + 2$ is negative this term 
will dominate the curvature term and we have a standard singularity ($V$ goes 
to minus infinity). This happens when 
\[
\frac{-6k(\gamma+1)+3\gamma+1}{3k(\gamma+1)-1} < 0.
\]
For dust matter it is satisfied if $k \in (-\infty,1/7) \cup (1/4, \infty)$.

In the opposite case if $k \in (0,1/7)$ we obtain a bounce as a scenario of 
an initial state. This scenario is the generic feature of 
cosmology inspired by quantum gravity effects \cite{Szydlowski:2005qb}.

Note that for $\gamma = 1/3$ effects of decaying $\lambda$ vanish and the 
model evolution follows the radiation dominating universe scenario. 

The motion of the system is defined on the zero energy level 
$H=\frac{\dot{a}^{2}}{2} + V(a) \equiv 0$. In the bouncing scenario the 
standard evolution with a singularity 
at $a = 0$ is replaced by a bounce. Note that negative contribution to 
$H^{2}(a)$ formula cannot dominate the material term and the Universe starts 
from some $a=a_{\text{min}} > 0$ --- a finite value of the scale factor. On 
the other hand, evolution can be prolonged into the domain $t < 0$ (a 
pre-bounce region) because of the reflectional symmetry $H \to -H$. If 
$\Omega_{K,0}<0$, i.e. the Universe is closed, then we obtain in the 
configuration space returning points. As a consequence we obtain 
characteristic type of evolution without initial singularity. 
Let us illustrate this situation. For the model with dust we have
\[
V(a) = - \frac{1}{2} \left[ \Omega_{K,0} + (1-\Omega_{K,0}) 
a^{\frac{-6k+1}{3k-1}} \right] \leq 0.
\]
This function has always zero: $a=a_{0} = \left( -\frac{\Omega_{K,0}}{1-\Omega_{K,0}} 
\right)^{\frac{3k-1}{-6k+1}}$ and $a \geq a_{0}$ for $k \in (1/6,1/3)$ and 
$a < a_{0}$ otherwise. The derivative of $V(a)$ determines a domain of 
acceleration
\[
\ddot{a} = - V'(a) = -\frac{1}{2} \frac{-6k+1}{3k-1} (1 - \Omega_{K,0}) 
a^{\frac{2-9k}{3k-1}}. 
\]
Therefore for any $t$ the scale factor acceleration $\ddot{a} > 0$ takes place 
if $V'(a) < 0$ i.e., $k \in (1/6, 1/3)$. In the opposite case V(a) is a 
growing function of $a$ ($V'(a)>0$) and $k \in (0,1/6) \cup (1/3,\infty)$. 
At $a=a_{0}$ $H=0$ because $\dot{a}^{2} = -2V$. If $(-6k+1)/(3k-1) > 0$ 
then $V(a_{0})=0$ and it is a bouncing point at $a_{0}$. In the opposite case 
$V(a=0)=-\infty$, i.e. there is a singularity at $a_{0}$. Summing up, in the 
acceleration regime $V$ is a decreasing function of $a$, $k \in (1/6, 1/3)$, 
and there is a bouncing point at $a_{0}$. 
In the opposite, deceleration regime $V$ is an increasing function of $a$,  
$k \in (0,1/6) \cup (1/3, \infty)$ and there is an initial singularity and 
a maximum size of the universe $a_{0}$. 

\begin{figure}
\includegraphics[width=0.8\textwidth]{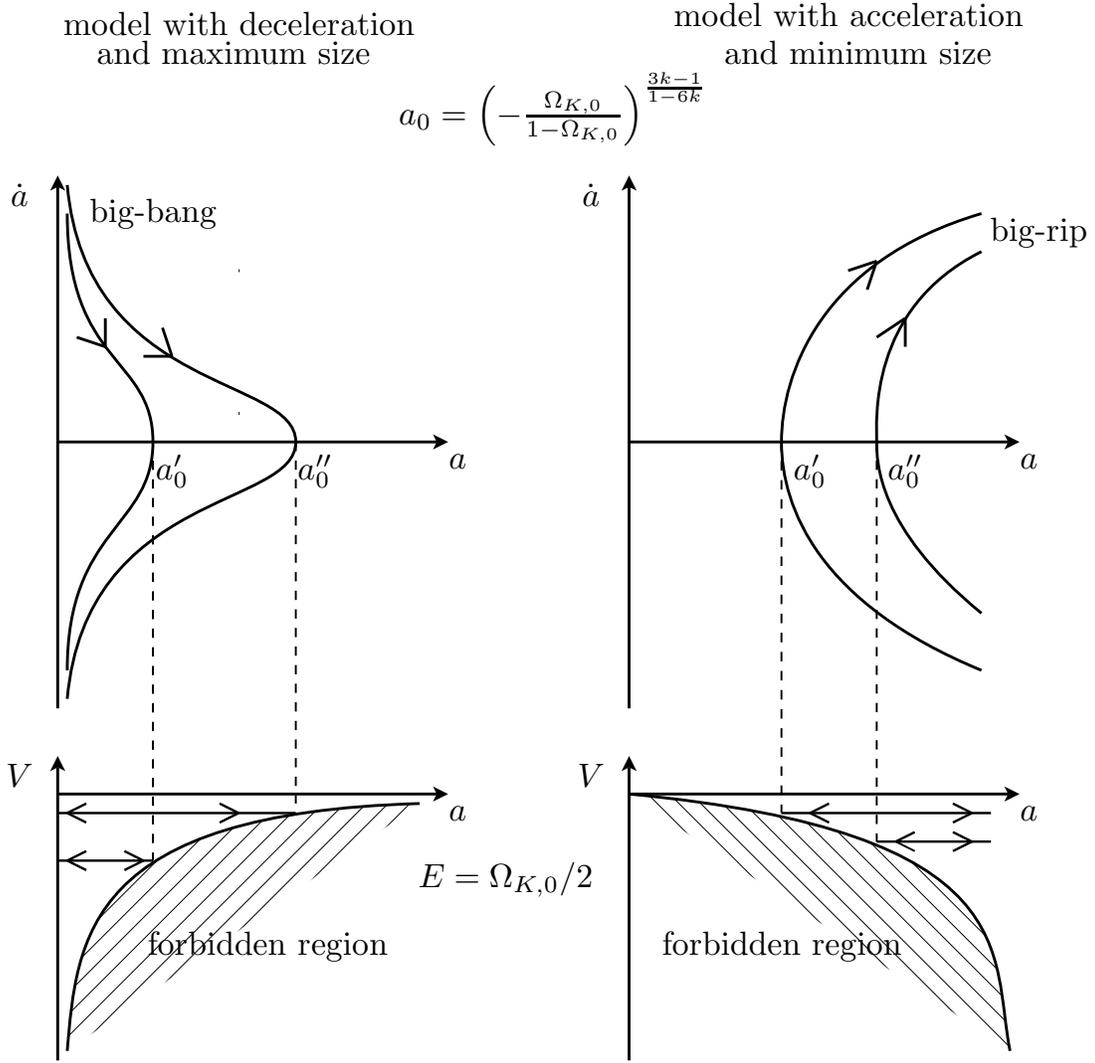}
\caption{The phase portrait and shape of diagram of potential function for 
Fomin model.}
\label{fig:4}
\end{figure}

The phase portrait of model for two cases is in Fig.~\ref{fig:4}.

In the literature there are well known propositions of explanation of 
acceleration of the Universe using cosmological models with modified 
Friedmann equations. Freese and Lewis \cite{Freese:2002sq} showed that the 
Cardassian modification of Friedmann equation (which contains a term of type 
$\Omega_{\text{\text{Card}},0} (1+z)^{3n}$, $n=\frac{4k-1}{3k-1}$) may give 
rise to acceleration without any dark energy of unknown origin contribution 
(see also \cite{Godlowski:2003pd} for the estimation of the model). 
In Fomin's model the age of the universe in the approximation of small 
$\Omega_{K,0}$ is given by formula
\[
T = \frac{2}{3} H_{0}^{-1} \frac{3k-1}{4k-1}.
\]
Therefore $k \simeq 1/6$ is required to explain the problem of age of the 
Universe.

\section{Conclusions}

We reviewed, clarified, and critically analyzed early approaches to Tryon's 
and Fomin's description of the origin of the Universe. We showed that the idea 
of the Universe's creation as a vacuum fluctuation event was discovered by 
them independently at around the same time. While Tryon's idea seems 
to be interesting only from the historical point of view, Fomin's approach 
is still heuristically fertile in description of latest issues in cosmology. 
Fomin's model brings the interesting explanation of the accelerating 
expansion of the Universe in terms of the Ricci scalar dark energy 
$\rho_{X} \propto R$ \cite{Gao:2007ep,Feng:2008rs,Kim:2008ej,Zhang:2009un}. 
This idea is important in the context of holographic principle of quantum 
gravity theory which regards black holes as the maximally entropic objects. 
For a given region of size $L$ effective field theory with UV cut-off $\Lambda$ 
gives the Bekenstein entropy bound $(L\Lambda)^3 \leq S_{\text{BH}}$ -- entropy 
of a black hole.

In this paper we study Fomin's idea of quantum cosmogenesis through the 
dynamical analysis. It is possible, in contrast to Tryon, because 
Fomin's conception is embedded in the environment 
of general relativity rather than in the Newtonian theory. We demonstrated 
that it is possible to construct the dynamical cosmology with incorporation of 
Fomin's mechanism of gravitational decaying of vacuum. This cosmology can be 
treated as a cosmology with varying cosmological constant. In Fomin' model 
appears the free parameter $k$. There are some interval of this parameter
for which the model accelerates and possesses a bounce instead of an 
initial singularity. It is necessary then to estimate it from the 
astronomical observations.

As recent supernovae of type Ia measurements indicate the expansion of our 
Universe is presently accelerating. While the cosmological constant offers 
the possibility of reconstruction of the effective theory of acceleration, 
the presence of fine-tuning difficulties motivates theorists to investigate 
alternative forms of dark energy. All these propositions can be divided into 
two groups. In the first group, unusual properties of dark matter with 
negative pressure violating the strong energy condition are postulated. In the 
other one, the modification of the Friedmann equation is postulated. In this 
approach, instead of a new hypothetical energy component of unknown form, some 
modifications of the FRW equations are proposed a priori (unfortunately 
without any fundamental justification).

The status of Fomin's cosmology seems to be similar to the status of the brane 
cosmological models where our observable Universe is a surface or a brane 
embedded in a higher dimensional bulk spacetime in which gravity could spread. 
Note, that this model with radiation can be recovered if parameter $k$ is $1/5$. 
We point out that Fomin's idea can be tested (in contrast to Tryon's idea) by 
using type Ia supernovae data. 

It is also important that Fomin's approach is free from the assumption of 
zero energy which is crucial for Tryon. Recently many authors concluded that 
energy-momentum tensor of the FRW universes are equal to zero locally and globally 
(flat universe) or just globally (closed universe). However, such a conclusion 
originated from coordinate-dependent calculations performed in the special 
comoving coordinates (Cartesian coordinates). 

It is interesting that predictions of this theory can be also tested by both 
Wilkinson Microwave Anisotropy Probe (WMAP) observations as well as by 
observations of distant supernovae type Ia (SNIa). We confront the Fomin FRW 
model with SNIa observation and estimate all parameters of models. Our analysis 
showed that Fomin's idea plays not only a historical role but is related to the 
modern cosmological models appeared in the context of dark energy.

The idea that sum of energies of all particles in the Universe is the same 
order of magnitude like gravitational energy, and consequently the total 
energy vanishes is very old. Overduin and Fahr \cite{Overduin:2003ve} argued 
that Haas \cite{Haas:1936ap} and Jordan \cite[p.16]{Jordan:1947hs} introduced this 
idea in 1936 and 1947, respectively, and called it the Haas-Jordan idea. The 
main difficulty in extending it to relativistic cosmology is caused by finding 
the correct definition of the energy in the FRW universe. The problem was not 
solve so far. On the other hand it was proposed to use the Newtonian framework 
to show this idea at work. But this approach basing on the concept of infinite 
$R^3$ space as a model of real space is inconsistent with homogeneous and 
isotropic static distribution of matter.

The FRW general relativistic cosmological models can be represented as a 
motion of a fictitious particle moving in the potential well 
\cite{Szydlowski:2003cf}. 
The shape of the potential function is determined by both matter content and 
curvature. In the simplest case the potential function is a function of the 
scale factor only. The balance energy relation $T+V=0$ corresponds to the 
Hamiltonian energy constraint $H=0$. Note that division on kinetic and 
potential parts of energy has purely conventional character. The system under 
consideration has a natural Lagrangian. From the potential function the 
curvature term can be extracted in such a way that motion of the system is 
restricted to the energy level $H = E = - \frac{K}{2}$. In the Newtonian 
cosmology the kinetic part is zero and the solution is admissible only for 
a closed universe because the gravity is attracting. Hence the problem of 
extension of the Haas-Jordan idea for the general relativistic case seems to 
be crucial for deeper understanding of the cosmogenesis mechanism.

The weakness of the Fomin model is that it does not specify what kinds of 
matter are created. If you have several components, you get only the overall 
balance equation, but not how the created matter is divided between the 
components. However the modern cosmological context of Ricci scalar dark 
energy suggests that created energy may corresponds to dark energy component. 

We obtain from observation that value of the basic parameter $k$ is $0.21$ and 
$\Omega_{\text{Card},0}= 1.03$ as the best fit. This means that formally 
fomin model is equivalent to the original Cardassian one with 
$\Omega_{K,0} = -0.03$. The value of density parameter of curvature 
is in good agreement with current CMB data \cite{Komatsu:2008hk}.
We also calculated the PDFs and confidence level of two parameters of 
Our analysis showed that in contrast to the Tryon model the Fomin model 
can be statistically estimated and in the light of SNIa data can explain 
accelerating phase of expansion of the current Universe.

From the philosophical point of view it means that there is some class of 
models of quantum cosmogenesis which are susceptible to falsify. 
Falsificability of Fomin's conception means that this theory becomes 
scientific in Popper's sense \cite{Popper:1959}.

\acknowledgments

The authors are very grateful to prof. M. Heller, dr A. Krawiec and A. Kurek 
for remarks and comments. Marek Szyd{\l}owski is especially very grateful to 
prof. Y. Shtanov for information on a detailed publication history of Fomin's 
paper and the fruitful discussion on quantum cosmogenesis. The authors also 
thank prof. J. Overduin and prof. A. Vilenkin for comments and historical 
remarks. This work was supported in part by the Marie Curie Actions Transfer 
of Knowledge project COCOS (contract MTKD-CT-2004-517186) and Polish State 
Committee for Scientific Research (KBN) grant No. 1 H01A 019 30.

\end{document}